\documentclass[%
 aip,
 amsmath,amssymb,
 reprint,%
]{revtex4-1}
\usepackage{CJK}
\usepackage{graphicx}
\usepackage{dcolumn}
\usepackage{bm}

\usepackage[utf8]{inputenc}
\usepackage[T1]{fontenc}
\usepackage{mathptmx}
\usepackage{etoolbox}

\makeatletter
\def\@email#1#2{%
 \endgroup
 \patchcmd{\titleblock@produce}
  {\frontmatter@RRAPformat}
  {\frontmatter@RRAPformat{\produce@RRAP{*#1\href{mailto:#2}{#2}}}\frontmatter@RRAPformat}
  {}{}
}%
\makeatother
\usepackage{fancyhdr}

\begin{document}

\preprint{AIP/123-QED}

\title{Vortex-Induced Drag Forecast for Cylinder in Non-uniform Inflow}

\author{Jiashun Guan (\begin{CJK*}{UTF8}{gbsn}关家顺\end{CJK*})}
 
\author{Haoyang Hu (\begin{CJK*}{UTF8}{gbsn}胡皓阳\end{CJK*})}%

\affiliation{ 
Department of Mechanics and Engineering Science, College of Engineering, Peking University, Beijing 100871, China
}%

\author{Tianfang Hao (\begin{CJK*}{UTF8}{gbsn}郝天放\end{CJK*})}
 
\affiliation{%
School of Electronics Engineering and Computer Science, Peking University, Beijing 100871, China
}%
\author{Huimin Wang (\begin{CJK*}{UTF8}{gbsn}王慧敏\end{CJK*})}
\affiliation{ 
Department of Mechanics and Engineering Science, College of Engineering, Peking University, Beijing 100871, China
}%

\author{Yunxiao Ren (\begin{CJK*}{UTF8}{gbsn}任耘霄\end{CJK*})*}
\affiliation{ 
Department of Mechanics and Engineering Science, College of Engineering, Peking University, Beijing 100871, China
}%
 \email{ renyx@pku.edu.cn}

\author{Dixia Fan (\begin{CJK*}{UTF8}{gbsn}范迪夏\end{CJK*})}
\affiliation{ 
School of Engineering, Westlake University, Hangzhou 310024, China
}%

\date{\today \quad Accepted by \textit{Physics of Fluids} (08 June 2025) }

\begin{abstract}
In this letter, a physics-based data-driven strategy is developed to predict vortex-induced drag on a circular cylinder under non-uniform inflow conditions -- a prevalent issue for engineering applications at moderate Reynolds numbers. Traditional pressure-signal-based models exhibit limitations due to complex vortex dynamics coupled with non-uniform inflow. To address this issue, a modified fully connected neural network (FCNN) architecture is established that integrates upstream velocity measurements (serving as an inflow calibration) with pressure-signal-based inputs to enhance predictive capability ($R^2\sim0 \rightarrow 0.75$). Direct numerical simulations (DNS) at Reynolds number $Re = 4000$ are implemented for model training and validation. Iterative optimizations are conducted to derive optimized input configurations of pressure sensor placements and velocity components at upstream locations. The optimized model achieves an $R^2$ score of 0.75 in forecasting high-amplitude drag coefficient fluctuations $(C_d=0.2$ -- $1.2)$ within a future time window of one time unit. An exponential scaling between model performance and optimized pressure signal inputs is observed, and the predictive capability of sparsely distributed but optimized sensors is interpreted by the scaling. The optimized sensor placements correspond to the physical mechanism that the flow separation dynamics play a governing role in vortex-induced drag generation. This work advances machine learning applications in fluid-structure interaction systems, offering a scalable strategy for forecasting statistics in turbulent flows under real-world engineering conditions.
\end{abstract}

\maketitle
The flow around cylindrical structures has attracted sustained research attention for over a hundred years \cite{zdravkovich1997,von1911,konstantinidis2011,norberg1987,williamson1996,lin2021,kravchenko2000}, motivated by their fundamental flow configuration and practical implementations encompassing ocean engineering systems, wind engineering applications, building aerodynamics, and so on. In real-world engineering applications, the non-uniform inflow conditions always introduce turbulence effects \cite{bearman1983}, exhibiting more complex characteristics relative to wake dynamics of cylinder flows \cite{williamson1996,bhattacharyya2022experimental}, e.g. the shrunk recirculation bubbles of the mean flow and high-amplitude fluctuations in drag (lift) coefficient $C_d$ ($C_l$) \cite{song2022}. 

\begin{figure}[!h]
    \centering
    \includegraphics[width=0.66\linewidth]{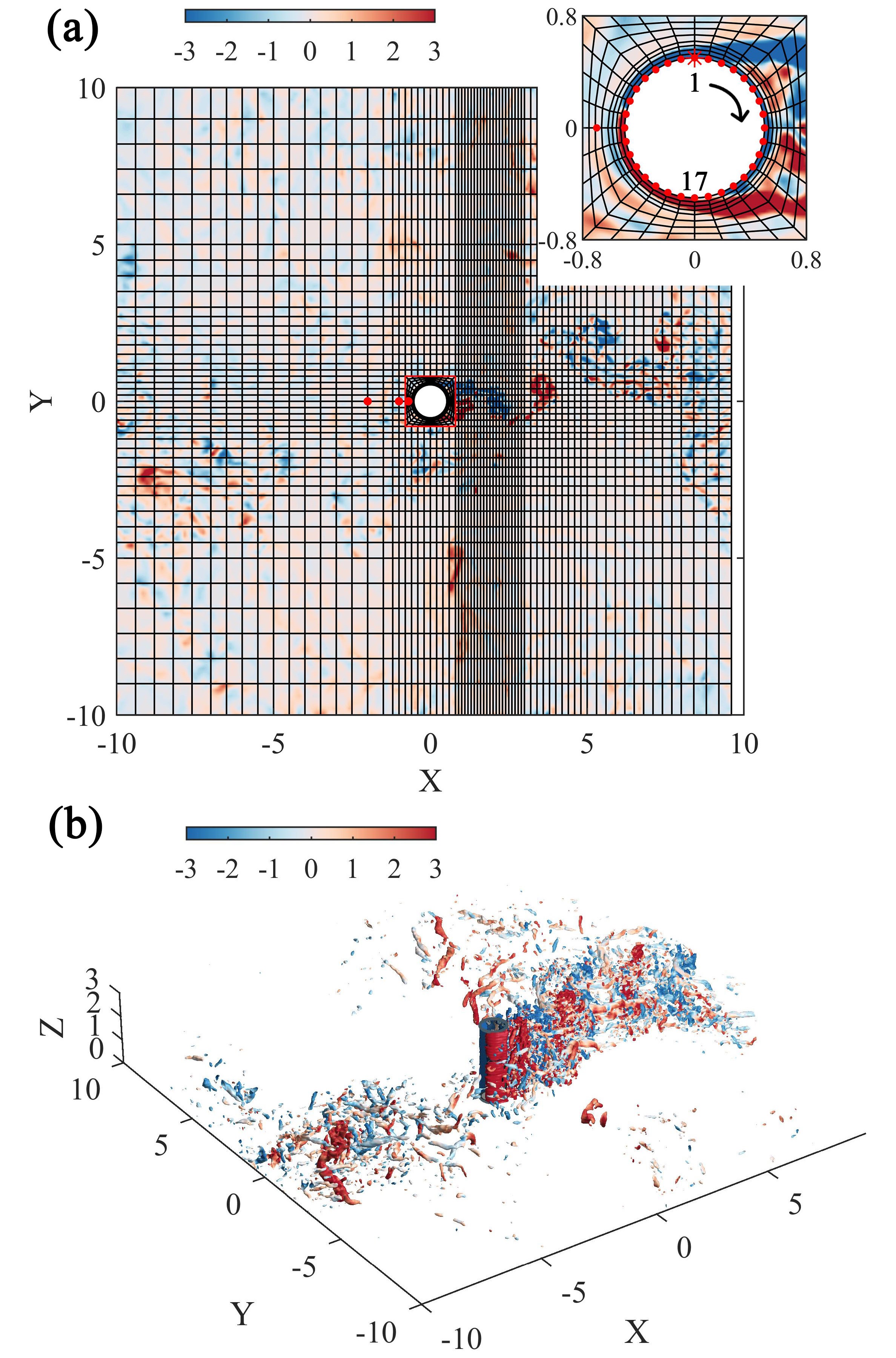}
    \caption{(a) The spatial distribution of spectral elements and the $Z$-component of vorticity in the $X-Y$ cross-sectional plane ($Z=1.5$). The insert is zoomed view of the red box, and red points indicates uniformly distributed locations of monitored points. The asterisk in the insert indicates the first monitored point ($pt=1$), with label number increasing clockwise around the circle.  (b) Visualization of turbulent structures through vortex criterion $Q=1.5 ~U^2/D^2$, colored by the $Z$-component of vorticity. }
    \label{fig:f1}
\end{figure}

To date, measuring \cite{bhattacharyya2023,guan2021,khan2023} and forecasting \cite{barthel2023,lee2019} dynamic evolution processes in fluid-structure interaction systems present technical challenges, while remaining critically important for preventing structure damage induced by extreme events \cite{sapsis2021} and frequency lock-in \cite{bearman1983,lin2021}. By leveraging machine learning and data-driven methods  \cite{fan2020,brenner2019,brunton2020}, progresses are achieved in areas such as statistics prediction, flow field reconstruction, flow control, and inverse problem resolution.  Barthel and Sapsis \cite{barthel2023} proposed a data-driven method for predicting airfoil extreme events using surface pressure signals processed by wavelet transformation, enabling accurate forecasting with a simple feed-forward neural network and minimal sensor deployment. Kim and Sapsis \cite{Kim2024} achieved real-time lift coefficient prediction for a dynamically stalling airfoil by employing discrete wavelet transforms to process surface pressure signals. 

\begin{figure*}
    \centering
    \includegraphics[width=0.95\linewidth]{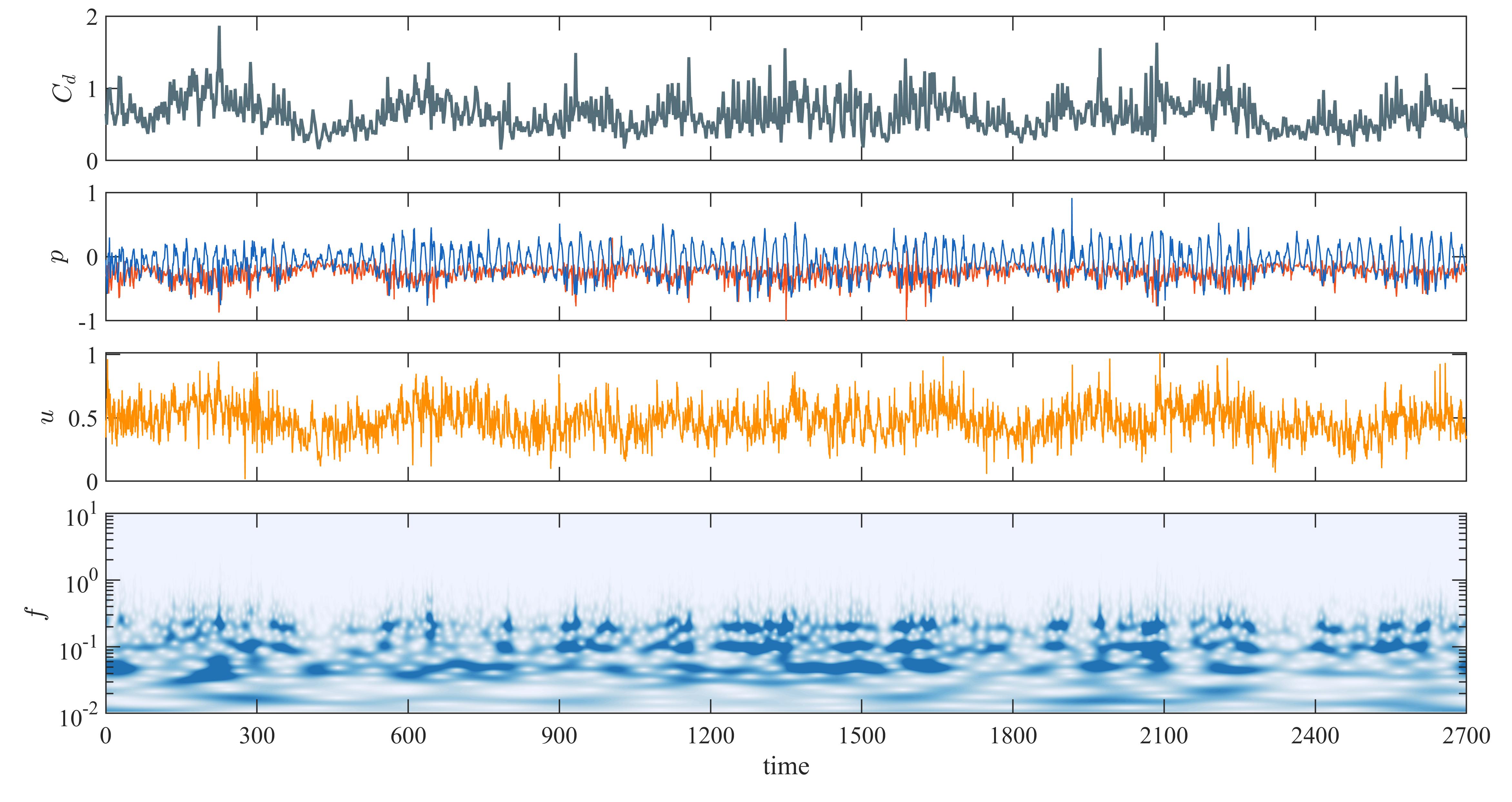}
    \caption{Under non-uniform inflow conditions, the following measurements are presented: drag coefficient $C_d$, pressures $p$ at $pt=4$ (blue) and $pt=21$ (red), streamwise velocity $u$ at $(-1,0,0)$, and continuous wavelet transform (CWT) processed $C_d$ signal. The contour in the bottom panel represents amplitudes of the frequencies $f$, linearly spanning the range $[0,0.1]$.}
    \label{fig:f2}
\end{figure*}

While the aforementioned works have demonstrated success in airfoil lift prediction through wavelet-processed pressure signals and neural networks, the pure pressure-signal-based model faces inherent limitations when applied to cylinder flow problems -- particularly for drag coefficient prediction under non-uniform inflow conditions that commonly characterize cylinder flows in real-world engineering applications at moderate Reynolds number regimes. In such cases, the complex vortex shedding dynamics, three-dimensional wake instabilities, and spatially heterogeneous pressure distributions induced by non-uniform inflows weaken the direct correlation between surface pressure signatures and drag force statistics. By incorporating physical interpretations to circumvent conventional brute-force testing approaches, this study therefore establishes a modified modeling framework specifically designed to be capable of these challenges.  

Direct numerical simulations (DNS) of flow over a circular cylinder are implemented at Reynolds number $Re=UD/\nu=4000$, where $U$, $D$ and $\nu$ are the flow rates, cylinder diameter and kinematic viscosity, respectively. The flow is governed by the incompressible Navier-Stokes equations:
\begin{subequations}
\begin{equation}
     \frac{\partial \mathbf{u}}{\partial t} + (\mathbf{u} \cdot \nabla) \mathbf{u}  = -\nabla p + \frac{1}{Re} \nabla^2 \mathbf{u}
    \label{eq:navier_stokes}
\end{equation}
\begin{equation}
    \nabla \cdot \mathbf{u} = 0
    \label{eq:continuity}
\end{equation}
\end{subequations}
where $\mathbf{u}$ is the vector of velocities $(u,v,w)$ and $p$ is the pressure field. The equations are nondimensionalized by setting the flow rates $U$ and the cylinder diameter $D$ as the characteristic velocity and length scales,respectively, and solved by the open source spectral element code Nek5000 \cite{nek5000_webpage_2008}. Periodic boundaries are applied in streamwise, spanwise, and cylinder-paralleled directions, respectively, leading to a non-uniform inflow of the cylinder. Spatial discretizations utilize 17020 spectral elements in $7^{th}$ order accuracy. The periodic cylinder is in $3 ~D$ length and discretized into $5$ elements. The streamwise-spanwise sectional view of spectral elements are shown in Fig.~\ref{fig:f1}a. The time stepping is set to $0.0005 ~D/U$, resulting in a CFL number of approximately 0.5 to ensure numerical stability.  The simulation spans a total dimensionless time of $3000~D/U$, comprising an initial transient phase ($300~D/U$) for flow development and a subsequent statistically stationary period ($2700~D/U$) under non-uniform inflow conditions for turbulence statistics acquisition. Velocities and pressures at 32 points along the cylinder and three upstream points located at $(-2,0,0)$, $(-1,0,0)$, and $(-0.7,0,0)$ (shown in Fig.~\ref{fig:f1}a), are dumped every $20$ time steps ($\Delta t=0.01~D/U$). The numerical setup is validated by comparing time-averaged ($1000 D/U$) pressure distributions under uniform inflow conditions against previous experimental results \cite{norberg1994}, demonstrating good agreement within $3.4\%$ deviation across the monitored point locations. 

\begin{figure}[t]
    \centering
    \includegraphics[width=0.95\linewidth]{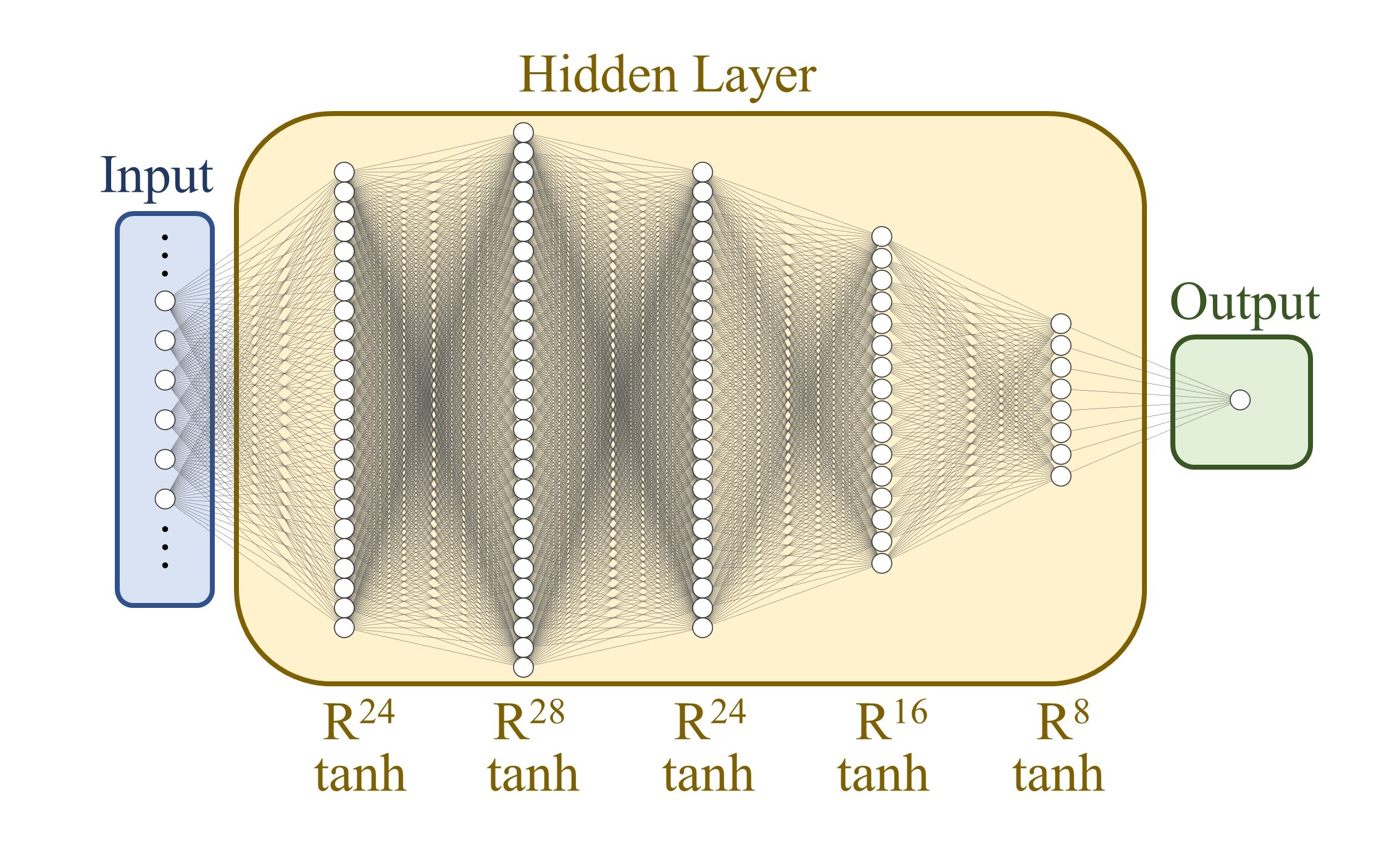}
    \caption{Architecture diagram of the fully connected neural network (FCNN), where $R^i$ represents the length of each fully connected layer.}
    \label{fig:fcnn}
\end{figure}

Numerical results in Fig.~\ref{fig:f2} are partitioned into training and testing sets ($2400 ~D/U$ and $300~D/U$, respectively) for subsequent $C_d$ forecasts using a fully connected neural network (FCNN), which employs the architecture: input-FC24-FC28-FC24-FC16-FC8-output (Fig.~\ref{fig:fcnn}). The output layer with fixed length $1$ is the predicted drag coefficient $C_d$ at a future time window of $\Delta t_p=1 ~D/U$ (unless otherwise specified in this study), while the hidden layer maintains a fixed architecture. Note that the input layer exhibits variable dimensions, and this study primarily aims to identify optimized input configurations. Each fully connected layer is followed by a hyperbolic tangent activation function $tanh$ to provide nonlinearity to the network. Moreover, to ensure consistent scale across input features, all training set samples were linearly normalized to the $[-1, 1]$ range using a min-max scaler. Training acceleration and overfitting mitigation were achieved through a cosine annealing learning rate scheduler with warm restarts coupled with the AdamW optimizer \cite{loshchilov2019}. The learning rate is initialized at $10^{-3}$ (truncated at $10^{-4}$) with a weight decay coefficient of $10^{-4}$. The training epoch is set to be $50$, and the warm-up epoch is $3$. Mean absolute error (MAE) is used as the loss function: 
\begin{equation}
    MAE=\frac{1}N{}\sum_{i=1}^{N} |\hat{C}_d(t)-C_d(t)|,
\end{equation}
where $\hat{C}_d(t)$ and $C_d(t)$ are the prediction and the training set in length $N$, respectively. The model performance is quantified via the $R^2$ score between predictions and DNS results in the testing set:
\begin{equation}
    R^2=1-\frac{1}{N}\sum_{i=1}^{N}\frac{(\hat{C}_d-C_d)^2}{\sigma^2},
\end{equation}
where $\sigma^2$ is the variance of $C_d$ in the testing set. In this study, $R^2$ is computed as the mean value obtained from five independent training trials of the identical model configuration.  

\begin{figure}[b]
    \centering
    \includegraphics[width=0.95\linewidth]{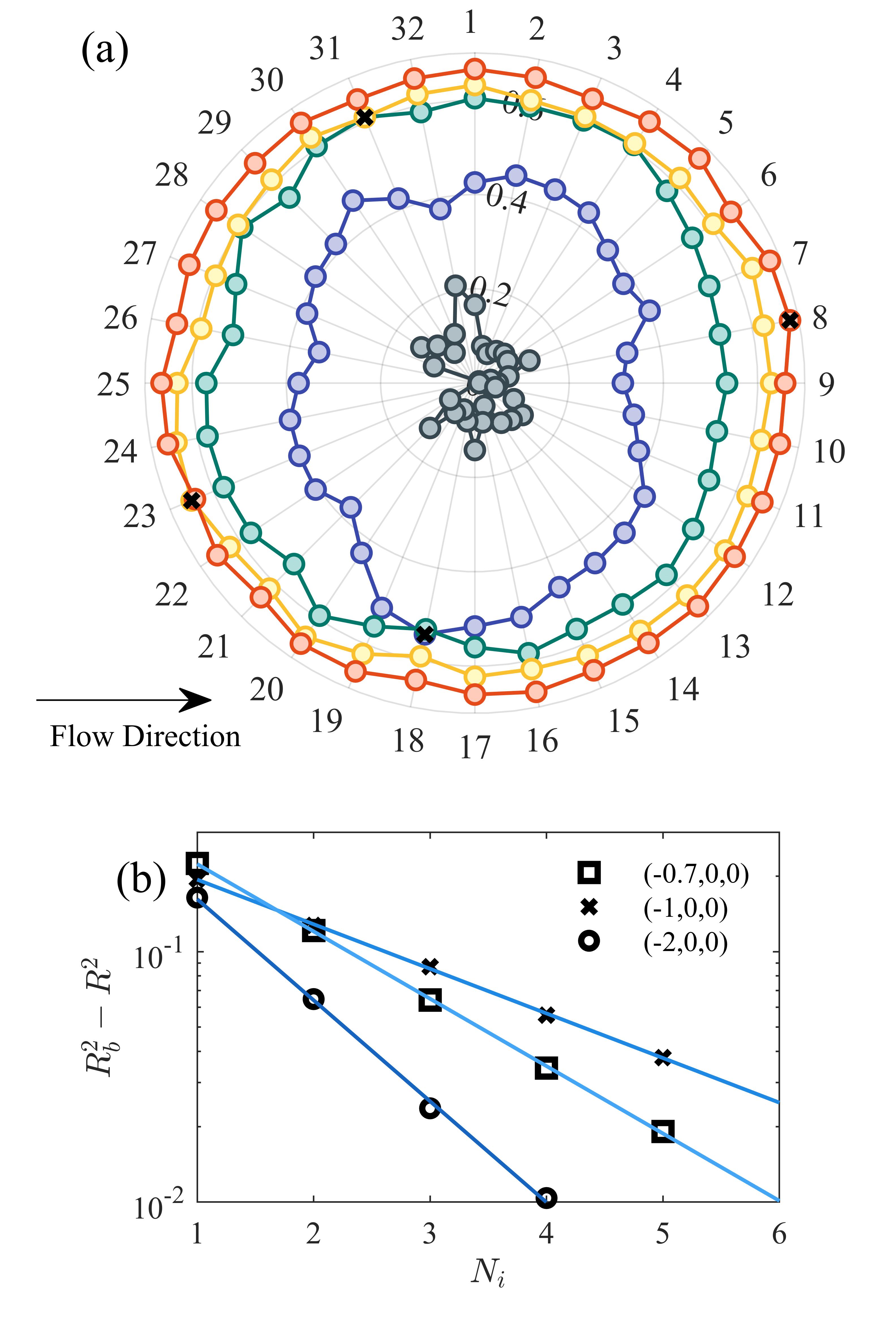}
    \caption{(a) The optimization iterations across 32 measurement points. The cross symbols $\times$ represent the maximum $R^2$ achieved during respective optimization iterations. The gray symbols are results of input with pure pressure signal ($\gamma$ and $d\gamma/dt$), while color symbols are obtained by input with modified $\gamma$, $d\gamma/dt$, $\mathbf{u}=(u,v,w)$ at $(-1,0,0)$, and $d\mathbf{u}/dt$. (b) The scaling laws (blue lines) of $R^2$ exponentially approaching the best performance $R^2_b$ with $N_i$, the number of optimized pressure signals in the input. The fitted $R^2_b$ are $0.65$, $0.74$, and $0.72$ with velocities at $(-0.7,0,0)$, $(-1,0,0)$, and $(-2,0,0)$, respectively. The cross symbols correspond to the results in (a).}
    \label{fig:f4}
\end{figure}

The measured drag coefficient $C_d$ was processed by the continuous wavelet transform (CWT) with the Morse wavelet \cite{olhede2012}, shown in Fig.~\ref{fig:f2}. Three distinct peaks in frequency space can be observed around $f=0.005$, $0.01$, and $0.02$. Based on CWT-processed results, a faster method, the discrete wavelet transform (DWT) \cite{sundararajan2015}, is applied to the preceding time window $\Delta t_w=2^n\Delta t$ of pressure signals, resulting amplitudes of the characteristic frequencies $\gamma_n(t)$, where the subscript $n$ denotes the $n^{th}$-level DWT coefficients. Following previously proposed path routes \cite{barthel2023,Kim2024}, the pressure signal at every single point is represented by a DWT-processed coefficient $\gamma(t)=\gamma_9(t)/2+\gamma_{10}(t)/2$ and its temporal derivative $d\gamma/dt\approx (\gamma(t)-\gamma(t-\Delta t))/\Delta t$, corresponding to the range of characteristic frequency peaks $f\in [0.005,0.02]$. The prediction models with inputs comprising single point pressure signal ($\gamma$ and $d\gamma/dt$) are trained respectively across $32$ points around the cylinder (gray symbols in Fig.~\ref{fig:f4}a). In contrast to the single-point-based prediction model under uniform inflow conditions \cite{Kim2024}, results in the non-uniform inflow conditions exhibit significantly degraded performance with $R^2\sim 0$. The underlying physical interpretation can be obtained straightforwardly: the non-uniform inflow conditions introduce multiscale frequency peaks (Fig.~\ref{fig:f2}), and the DWT-processed coefficient $\gamma$ fails to resolve the broadband spectral components.    

To expand the DWT-processed forecast algorithm into non-uniform inflow conditions, we add the velocity signals at the upstream point into the input array. The velocity signals provide a macroscopic calibration of inflow, substituting the low-frequency peaks in the dynamics of drag coefficient $C_d$. Therefore, the DWT-processed coefficient $\gamma$ of pressure signal is adjusted to $\gamma=\gamma_9(t)$, representing the high-frequency peaks $f\in [0.01,0.02]$, and the input for model training consists of velocities $\mathbf{u}=(u,v,w)$, their time derivations $d\mathbf{u}/dt\approx (\mathbf{u}(t)-\mathbf{u}(t-\Delta t))/\Delta t$,  DWT-processed coefficient $\gamma$, and its temporal derivative $d\gamma/dt$. Preliminary trials with velocities $\mathbf{u}=(u,v,w)$ at $(-1,0,0)$ and one pressure signal are carried out (purple symbols in Fig.~\ref{fig:f4}a), demonstrating measurable performance enhancements. The optimized combination of velocity and pressure signals will be illustrated in the following. 

Despite the low computational cost of model training,  the input space encompasses $2^{32+3\times 3}\sim 10^{12}$ distinct velocity-pressure signal combinations, and the global optimum is almost impossible to achieve. Therefore, velocity components $(u,v,w)$ in the input are provisionally fixed at point $(-1,0,0)$ to optimize pressure sensor placement. The pressure signal exhibiting maximum $R^2$ values is sequentially selected through iterative evaluation across all 32 measurement points along the circle (Fig.~\ref{fig:f4}a), and is added to the input for the next iteration. The optimizing iterations lead to an increasing $N_i$, which represents the number of optimized pressure signals in the input (Fig.~\ref{fig:f4}b). To constrain input array size, the optimizing iterations will be terminated when $N_i$ reaches $5$ or $R^2$ nears $99\%$ $R_b^2$. A scaling is observed that the max $R^2$ exponentially approaches its best performance $R^2_b$ when the optimized input containing the increasing $N_i$ (cobalt blue lines in Fig.~\ref{fig:f4}b). The best performance $R^2_b=0.74$ is obtained by fitting with the least squares method. The exponential scalings are also validated by the inputs with velocities at $(-2,0,0)$ and $(-0.7,0,0)$, shown in Fig~\ref{fig:f4}b by the light blue lines, respectively. The exponentially asymptotic trend indicates an efficient configuration that just several signal sensors can perform well enough on the $C_d$ forecast, e.g., the input with two pressure signals and velocities at $(-2,0,0)$ achieves performance better than $90\%~R^2_b$.  

\begin{figure}[h]
    \centering
    \includegraphics[width=0.95\linewidth]{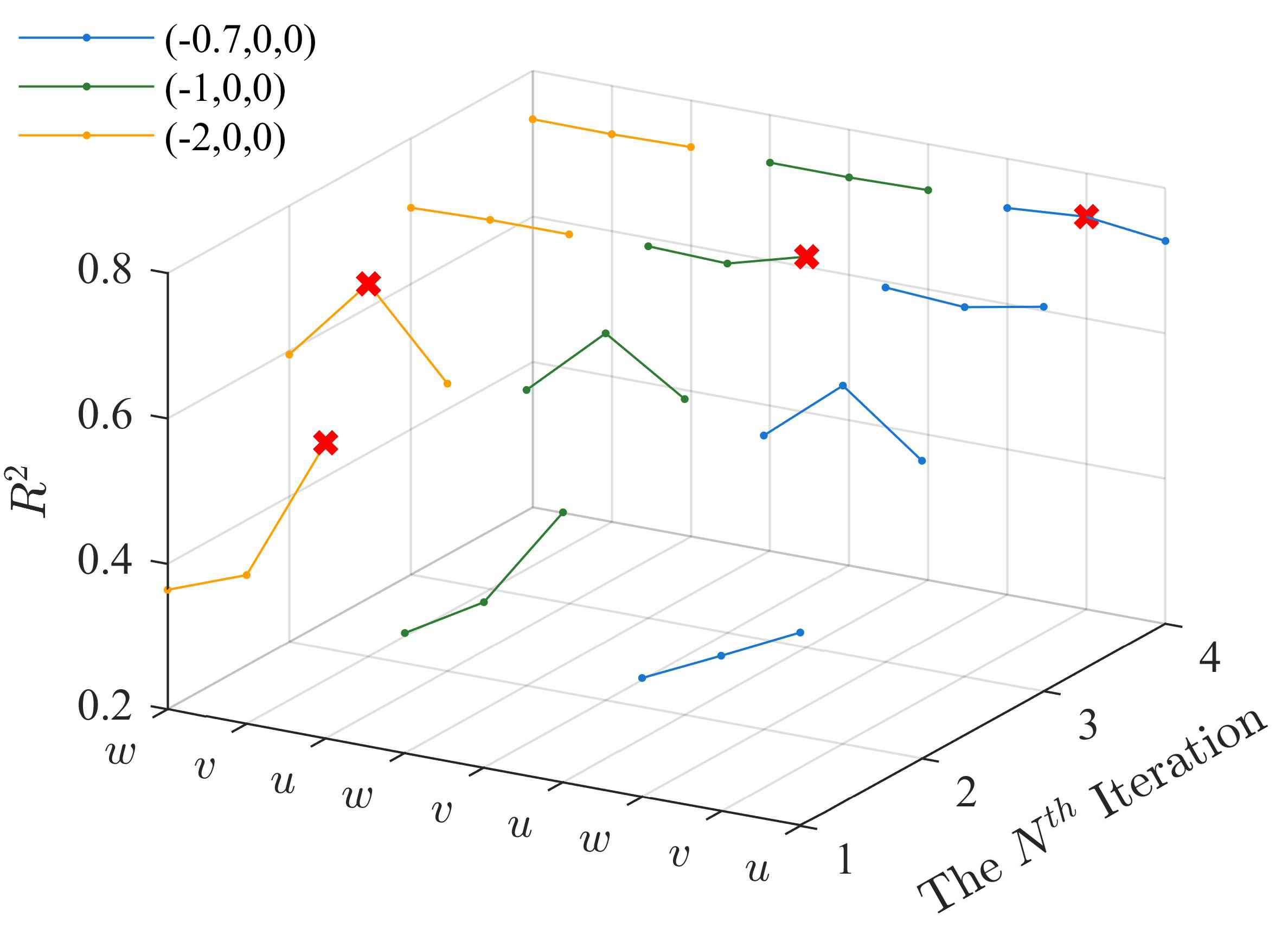}
    \caption{Optimization of the velocity combinations in the input. The red crosses $(\times)$ represent the max $R^2$ value in every iteration. The coordinates in the legend indicates the locations of velocity signals used in the model. }
    \label{fig:f5}
\end{figure}

The first four optimized pressure sensor locations are $pt= 18,30,8,22$,  $pt= 18,31,23,8$, and  $pt= 18,1,31,23$, acquired by inputs with velocities at $(-2,0,0)$, $(-1,0,0)$, and $(-0.7,0,0)$, respectively. Contrary to initial intuition, velocities acquired at a finite distance ($1$ -- $2~D$) upstream demonstrate superior forecast  capabilities ($R^2_b > 0.7$) compared to the near-wall counterpart ($R^2_b = 0.65$), and the following discussion will focus on the better ones. We found that the first two optimized $pt$ maintain spatial consistency across inputs with varying upstream velocities, and locate at the leading edges of flow separation points \cite{song2022}. This spatial correlation underscores the governing role of flow separation dynamics in vortex-induced drag generation mechanisms \cite{williamson1996,bloor1964,khabbouchi2014}. The third and fourth optimized $pt$ locate at the front leading edge ($pt=22, 23$) and trailing edge ($pt=8$) of the cylinder, indicating that hydrodynamic interactions in the corresponding flow region play a supplementary role for the forecast model. Similar results are also reported in the lift forecast of a airfoil \cite{Kim2024}.  The trailing edge region (particularly near the rear point, e.g. $pt9$) undergoes intermittent vortex shedding, enhancing fluctuations of the chaotic and multiscale pressure signals (Fig.~\ref{fig:f2}). This nonlinear dynamics induces informative leakage in its DWT-processed coefficient $\gamma$ and the temporal derivative $d\gamma/dt$, especially where the latter is in the first-order temporal accuracy. 

\begin{figure}[b]
    \centering
    \includegraphics[width=0.9\linewidth]{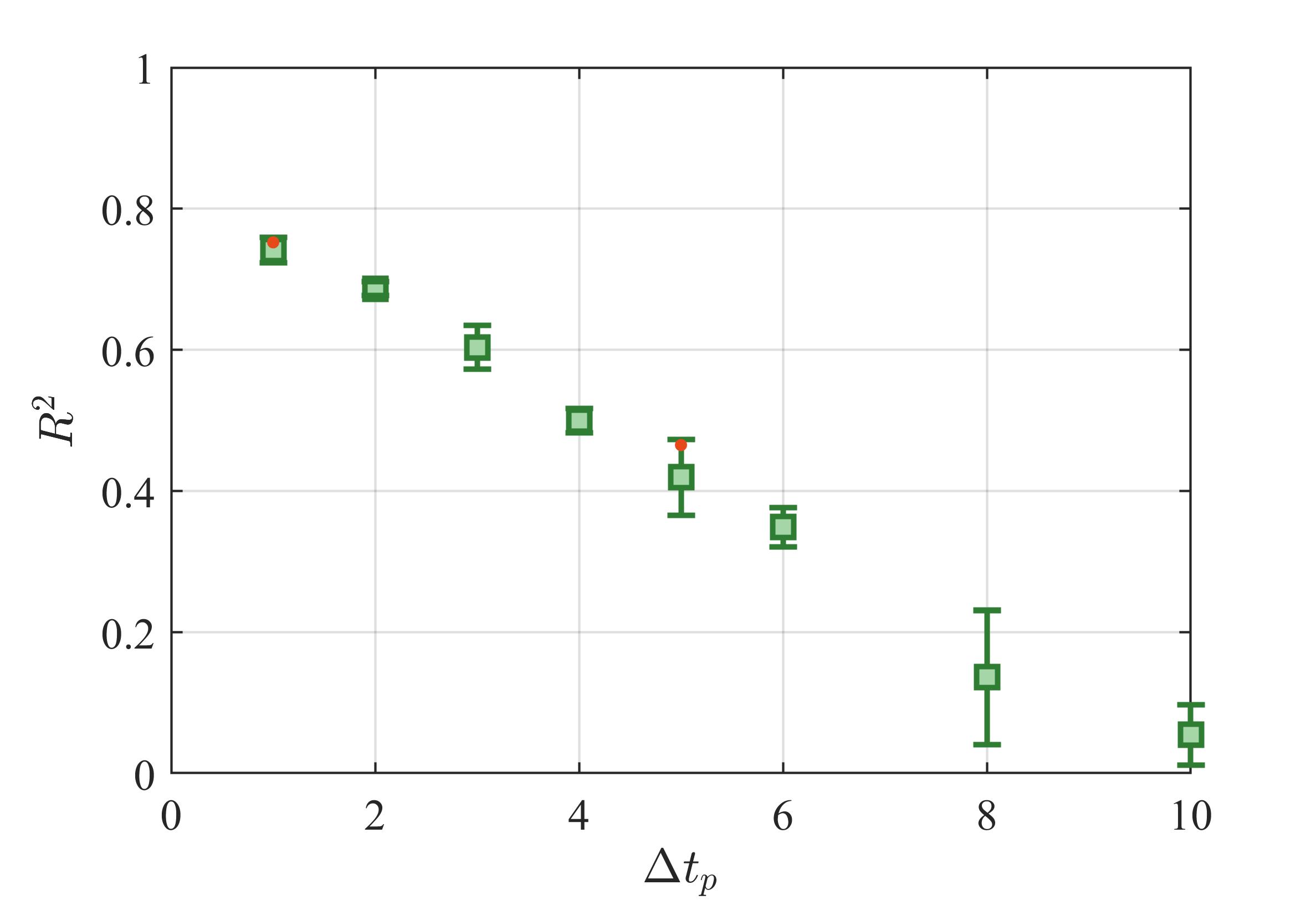}
    \caption{Prediction accuracy ($R^2$) in different future time window $\Delta t_p$. Error bars indicate $\pm 2$ standard deviations calculated from five independent training trials. The red points are two of independent training trials shown in Fig.~\ref{fig:f6}.}
    \label{fig:f7}
\end{figure}

\begin{figure*}
    \centering
    \includegraphics[width=0.9\linewidth]{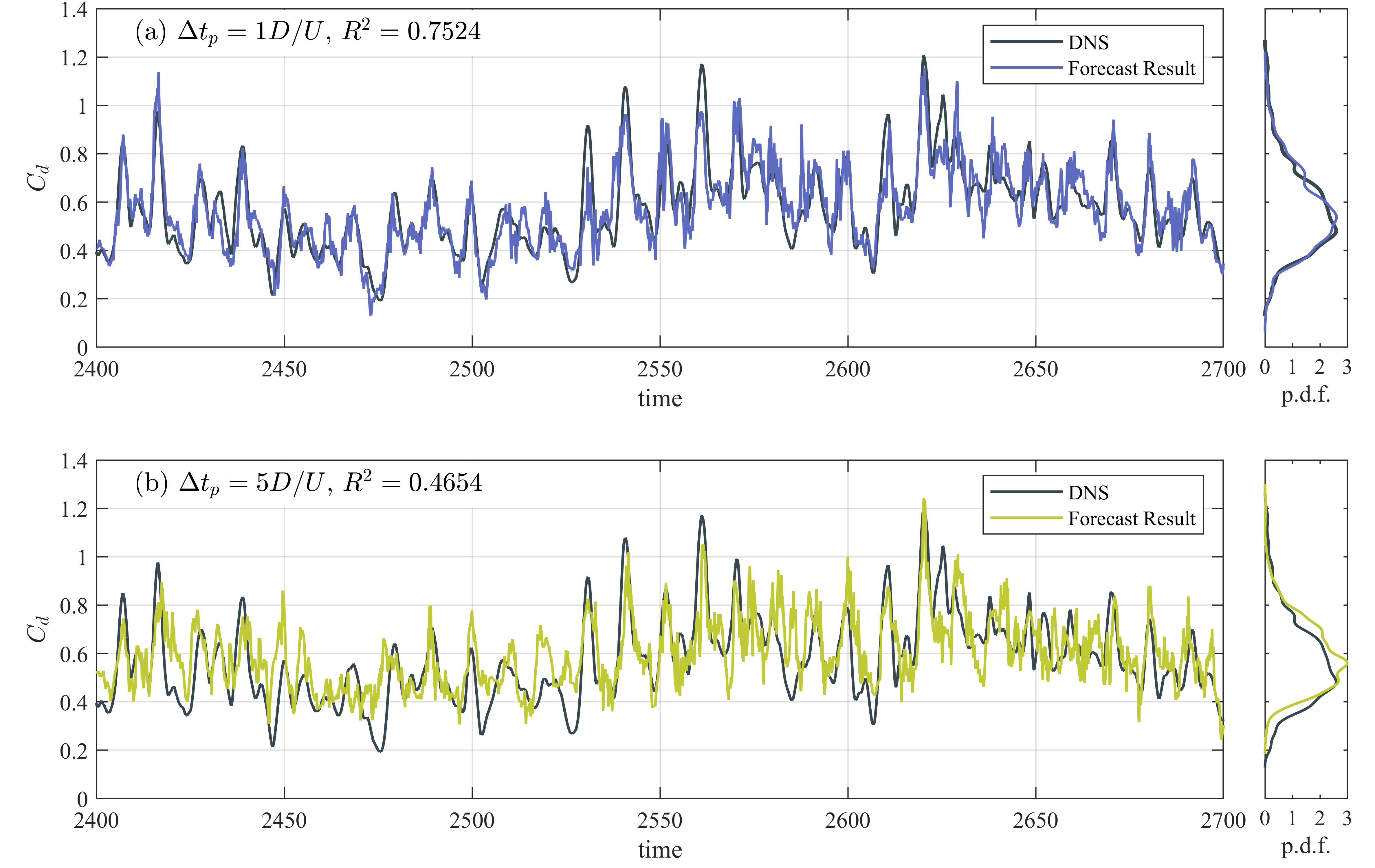}
    \caption{Drag coefficient $C_d$ in $\Delta t_p=1~D/U$ (a) and $\Delta t_p=5~D/U$ (b) predicted by the FCNN with optimized input are compared with the DNS benchmarks. The right-hand panels are comparisons of the possibility density functions (p.d.f.) between forecast results and the DNS benchmarks. }
    \label{fig:f6}
\end{figure*}

After obtaining the optimized pressure signal combination, velocities $(u,v,w)$ at three upstream points are decoupled into $9$ independent variables to optimize velocity combinations through iterative processes. Pressure sensor placements are specified as $pt= 18,30,8,22$ in this optimization, based on their more rapid convergence in the exponential scaling shown in Fig.~\ref{fig:f4}b. The results shown in Fig.~\ref{fig:f5} reveal several discoveries: (1) Upstream velocity component $u$ exhibits the highest information content, followed by $v$, generating a characteristic hat-shaped ($~\hat{}~$)  during the second iteration. (2) The $w$-component velocities show negligible contribution, consistent with the physical interpretation that the cylinder-paralleled $w$ components have limited influence on the inflow calibration. (3) Similar to the results of pressure sensor placement optimization, using as few as two velocity signals achieves significant improvement in prediction accuracy $R^2$ significantly ($R^2:~\sim0~\rightarrow ~\sim0.7$); thereby demonstrating the effectiveness of inflow calibration strategy. According to the results shown in Fig.~\ref{fig:f5}, the optimized input configuration comprises velocity signals at $u(-2,0,0)$, $v(-2,0,0)$, $u(-1,0,0)$, and $v(-0.7,0,0)$, combined with pressure signals $pt=18,30,8,22$. This combination achieves a forecast model with $R^2=0.75$, as demonstrated in Fig.~\ref{fig:f7} and Fig.~\ref{fig:f6}a. The curves of forecast results and DNS results nearly overlap across all temporal evolutions except that the forecast results show small-scale fluctuations. As the future time window $\Delta t_p$ increases, the model's predictive accuracy decreases monotonically (Fig.~\ref{fig:f7}). Note that the forecast results of fluctuating drag coefficient $C_d \sim0.2$ -- $1.2$ in the non-uniform inflow still maintain qualitative consistency with the DNS benchmarks, even under extended $\Delta t_p=5~D/U$ conditions (Fig.~\ref{fig:f6}b). 

The FCNN model developed in this study demonstrates sufficient accuracy while maintaining high computational efficiency, enabling real-time drag forecasting for a cylinder under complex non-uniform inflow conditions. To address the prediction inaccuracy caused by the non-uniform inflow, we have introduced upstream velocity measurements into the network input, which is not obtained by brute-force testings, but follows a physical interpretation that upstream velocities act as an inflow calibration. Therefore, a strategy combining optimized pressure signals along the cylinder surface and optimized upstream velocity signals is proposed and validated through comparison with DNS results. An exponential scaling emerges between $R^2$ and the number of optimized pressure signals incorporated in the input, illustrating that the first two optimized pressure signals exert a dominant influence on forecasting efficacy. Moreover, these optimized pressure signals robustly locate leading edges of the flow separation points, whose importance in the vortex-induced dynamics has been substantially demonstrated in previous literature \cite{williamson1996,bloor1964,khabbouchi2014,song2022}. The optimized results can be interpreted physically that slight leading-edge fluctuations can be nonlinearly amplified, ultimately governing the chaotic vortex dynamics in the downstream flow field. While the exponential scaling demonstrates robust performance across all investigated cases, its theoretical interpretation based on the Navier-Stokes equations remains elusive. Furthermore, we anticipate that the proposed inflow calibration strategy for forecasting statistics in complex flow environments \cite{song2022,bhattacharyya2023experimental,khabbouchi2014} holds potentials for experimental validation, flow control schemes, and engineering implementation.    

\section*{ACKNOWLEDGMENTS}
The authors are grateful for the support from the National Natural Science Foundation of China (Grant No. U24A20266). Computational resources are supported by superserver@js. J. Guan acknowledges insightful discussions with D. Kim at the \textit{26th International Conference of the Theoretical and Applied Mechanics} (ICTAM 2024).

\section*{DATA AVAILABILITY}
The data and code that support the findings in this study are available from the corresponding author upon reasonable request.

\bibliography{aipsamp}

\end{document}